# Flexible bandwidth 448 Gb/s DMT Transmission for Next Generation Data Center Inter-Connects

Annika Dochhan[(1)], Helmut Grießer[(2)], Michael Eiselt[(1)], Jörg-Peter Elbers[(2)]

[(1)] ADVA Optical Networking SE, Märzenquelle 1-3, 98617 Meiningen, Germany,
adochhan@advaoptical.com
[(2)] ADVA Optical Networking SE, Fraunhoferstr. 9a, 82152 Martinsried, Germany

**Abstract** *We experimentally evaluate a flexible DMT system using 4 to 8 50-GHz-grid C-band channels to transmit 448 Gb/s over up to 240 km SSMF. VSB filtering enabled by detuned lasers significantly reduces the impact of chromatic dispersion.*

**Introduction**
Discrete multitone transmission (DMT) as a special version of orthogonal frequency multiplexing (OFDM) recently attracted attention for short-reach interconnects with intensity modulation and direct detection (IM/DD).[1] Up to 100 Gb/s on a single C-band wavelength over 10 km of standard single mode fiber (SSMF) have been reported, overcoming the chromatic dispersion (CD) induced power fading effect in an OFDM IM/DD system by bit and power loading (BL/PL).[2] However, recent 100 Gb/s per wavelength DMT research focusses on transmission in the O-band for client side optics.[1,3] In our targeted application, the inter-connection between data centers, or short reach metro connections up to 200 km, the use of dense wavelength division multiplexing (DWDM) and optical amplification by erbium doped fiber amplifiers (EDFAs) are mandatory, so the use of the O-band is not an option. In earlier work, we investigated DMT at different data rates and transmission over up to 80 km of SSMF, showing that the performance significantly improves for data rates below 112 Gb/s.[4] The effect of CD can be mitigated by using a single sideband signal (SSB) and actually longest transmission reaches for optical DMT were achieved for a SSB version with a significant frequency gap around the optical carrier.[5] Such modulation concepts, however, require more complex optical components and RF circuitry. Nevertheless, some improvement is possible by asymmetrical filtering without requiring additional components, leading to a vestigial sideband (VSB) signal. Moreover, VSB filtering reduces the required optical bandwidth for each channel.[6] So far our experiments only covered single channel evaluations. Here, we propose a DWDM system with 50-GHz channel grid using a variable number of channels, depending on the desired transmission reach. Starting with five channels of 89.6 Gb/s each and a reach of 40 km, the adaptive 448 Gb/s system can also cover 240 km with eight channels of 56 Gb/s.

**Experimental setup**
Fig. 1 shows the experimental setup. All DMT building blocks were realized in python software and offline processing. At the transmitter we used two channels of a 64 GS/s analog-to-digital converter (DAC), which drive two 40G Mach-Zehnder modulators (MZM). Both channels carried the same signal, but cyclically shifted by a random number of samples. Each MZM modulates a group of four 100-GHz spaced wavelengths combined by polarization maintaining 1x4 couplers and with a relative offset of 50 GHz between the two groups. The driving voltages to the MZMs were around 20% of the switching voltage with a bias slightly above the electrical field's zero point to ensure linear modulation. The signal was clipped with a clipping ratio of 9-11 dB (depending on the data rate) to reduce the peak-to-average power ratio and to avoid negative values for the electrical field. After the MZM, the two modulated four-channel groups are combined by a 50-GHz interleaver (IL), leading to an eight-channel, 50-GHz spaced WDM signal. Depending on the data rate of each individual wavelength, edge wavelengths are switched off to fix the data rate at 448 Gb/s. Four to eight wavelengths require rates of 112 Gb/s, 89.6 Gb/s, 74.7 Gb/s, 64 Gb/s, and 56 Gb/s per wavelength, respectively. The signal was amplified and transmitted over up to three spans of 80 km SSMF. Each span shows a loss of ~18 dB. For the evaluation of shorter distances a single span with variable length (10 km, 20 km, 40 km) was also considered. After the transmission line, a variable optical attenuator (VOA) allowed the adjustment of the optical signal-to-noise ratio (OSNR) for single channel performance evaluations. The signal was then pre-amplified, even and odd channels were separated by another IL and the channels to be evaluated were filtered out by an optical demultiplexer. The filtered signals were amplified to ensure sufficient input power to the PIN photo diodes. The electrical signals were captured by an 80-GS/s real-time (RT) oscilloscope with an electrical bandwidth of 29.4 GHz. Two wavelengths, one of each group, were detected simultaneously.

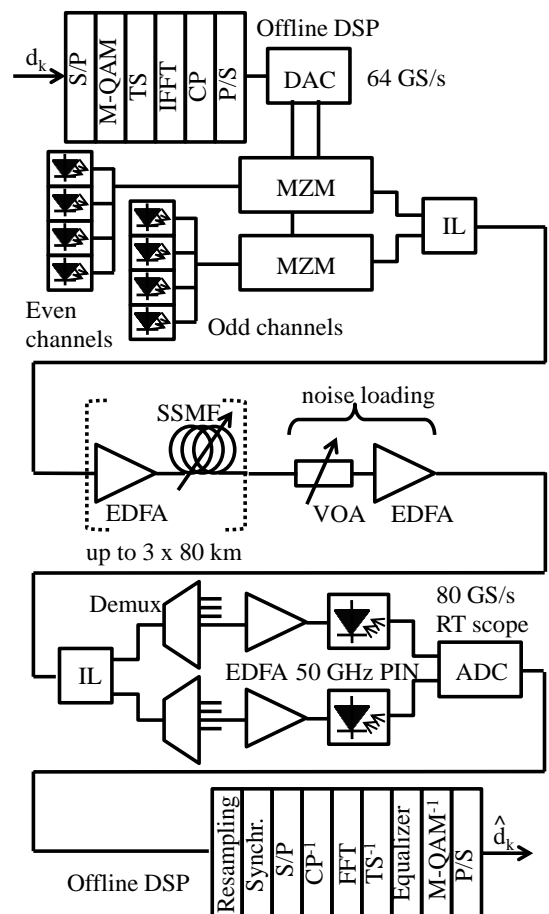

**Fig. 1:** Setup of the 8-channel 448 Gb/s DMT system.

For the generation of the DMT signal, a 2048-point FFT was used with up to 974 subcarriers carrying data to ensure a real-time baseband signal and an oversampling of 1.05. To properly assign the data to the subcarriers, the SNR was estimated by evaluating a transmitted training frame with uniform bit and power distribution. In a second step, BL and PL were performed according to Chow's margin adaptive algorithm.[7] One DMT frame consisted of 119 data symbols and 5 training symbols (TS) for channel estimation and synchronization. A cyclic prefix (CP) of 1/64 of the symbol duration was used. All data rates given here are net rates without TS and CP (but including some overhead for forward error correction, FEC). The detected signal was linearized by taking the square root. Resampling was followed by Schmidl-Cox synchronization.[8] After FFT and overhead removal, a decision directed one-tap equalizer was used to recover the signal, which could then be de-mapped for error counting.

**Single Channel Results and Discussion**

In a first step, we evaluated the influence of CD and filtering by the IL on a single channel signal. We stepwise increased the transmission length and measured the required OSNR for a BER of $4 \cdot 10^{-3}$, which can be considered as the limit for a good hard decision FEC to ensure error free transmission.[9] If the wavelength is centered in the IL passband, it will fully experience the CD induced power fading. In contrast, if the wavelength is shifted to the edge of the IL filter, it experiences more non-linear subcarrier-to-subcarrier interference after detection. Consequently, a trade-off has to be found. We optimized the frequency offset by introducing 50 km of SSMF, fixing the OSNR to 41 dB and measuring the BER for stepwise detuned signals. The optimum detuning turned out to be independent of the data rate. Since in our previous experiments[6] the optimum detuning did not change significantly with the transmission reach, we fixed the detuning to 19 GHz for all fiber lengths. Fig. 2 shows for different data rates the required OSNR vs. transmission reach for the centrally aligned signal and for the signal detuned by 19 GHz. The benefit of detuning can be clearly observed. Even 112 Gb/s can be transmitted on a single wavelength over up to 40 km. However, the high OSNR requirement for this data rate shows that it is the limit of what is possible with this system.

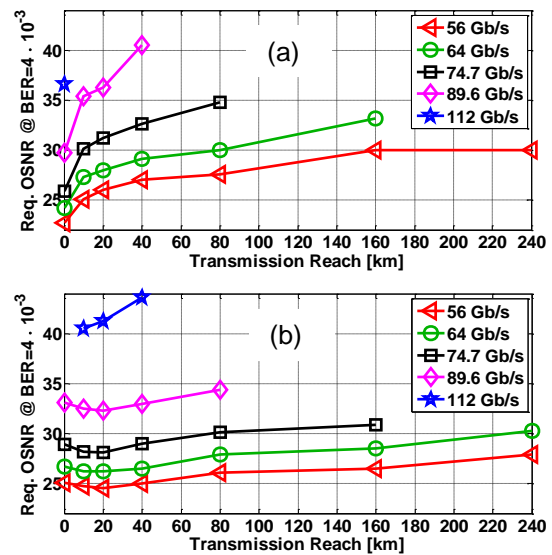

**Fig. 2:** Required OSNR vs. transmission reach for all considered data rates and (a) signal centered to the IL passband, (b) signal detuned by 19 GHz.

**WDM Transmission Results and Discussion**

In the next step, we transmitted all WDM channels. Back-to-back results already showed that linear crosstalk comes into play – the minimum achievable BERs were reduced for all data rates. Fig. 3 shows these BER values for all channels and data rates. 112 Gb/s does not achieve the FEC limit any more. The goal for the transmission experiments is to use as few WDM channels as possible to transmit the 448 Gb/s. So, for each fiber length we evaluated the maximum possible data rate per channel for which all channels achieve a BER below $4 \cdot 10^{-3}$. For this rate, we optimized the fiber launch power by measuring the BER for one center channel. Fig. 4 shows the BER vs. launch power for different lengths and rates. The launch powers which were finally used for transmission were between 5 and 6 dBm per wavelength and are marked with solid symbols.

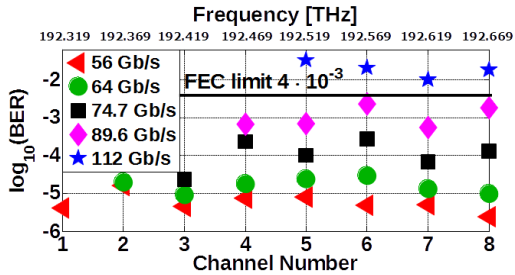

**Fig. 3:** Minimum achievable BER for all channels within the 448 Gb/s system in the back-to-back case.

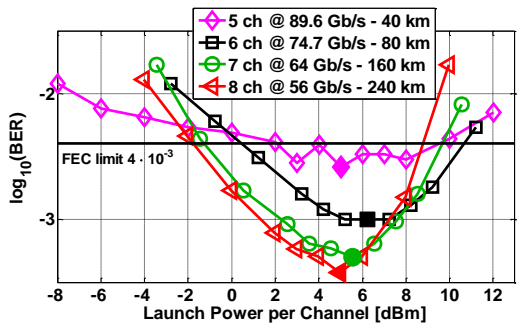

**Fig. 4:** BER vs. fiber launch power for the 448 Gb/s system for each channel count/data rate at maximum transmission reach. Solid markers indicate the optimum power.

It turns out that five channels with a rate of 89.6 Gb/s could reach 40 km of SSMF, while six channels with 74.7 Gb/s could be transmitted over 80 km. With seven 64-Gb/s channels the maximum reach was 160 km, and with eight channels of 56 Gb/s even 240 km could be achieved. Fig. 5 demonstrates that for these reach/data rate/channel count scenarios the BER of all wavelengths was below the FEC limit.

## Conclusion
We propose a 448 Gb/s DMT system targeting short-reach metro and data center interconnections. Dependent on the required reach, the number of DMT channels and thus the total bandwidth can be chosen flexibly. By detuning the laser frequency, the interleaver filters realizing the 50-GHz channel grid can be used as vestigial sideband filters to mitigate the chromatic dispersion induced power fading, which is inherent to DD DMT systems.

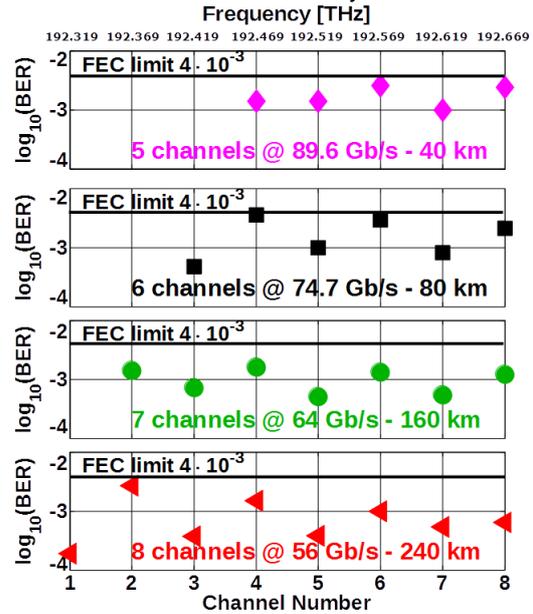

**Fig. 5:** BER after transmission for all channel count/data rate/reach scenarios of the 448 Gb/s DMT system.

## Acknowledgements
This work was funded by the German ministry of education and research (BMBF) within the SASER ADVAntage-NET project under contract 16BP12400.